\documentclass[12pt,pra,aps,amssymb,amsfonts,amsmath,tightenlines]{revtex4}
\usepackage{dsfont}
\usepackage[dvips]{graphicx}
\usepackage{amssymb,amsfonts}

\newcommand{\half}{\mbox{$\textstyle \frac{1}{2}$}}

\newcommand{\re}{\mbox{$\rm e$}}
\newcommand{\ri}{\mbox{$\rm i$}}
\newcommand{\rd}{\mbox{$\rm d$}}

\begin{document}

\title{Credit Risk, Market Sentiment and Randomly-Timed Default}

\author{Dorje~C.~Brody${}^1$, Lane~P.~Hughston${}^1$, and 
Andrea Macrina${}^2$}

\affiliation{${}^1$Department of Mathematics, Imperial College
London, London SW7 2BZ, UK \\ ${}^2$Department of Mathematics,
King's College London, London WC2R 2LS, UK, and \\ Institute of 
Economic Research, Kyoto University, Kyoto 606-8501, Japan}


\begin{abstract}
We propose a model for the credit markets in which the
random default times of bonds are assumed to be given as functions
of one or more independent ``market factors". Market participants
are assumed to have partial information about each of the market
factors, represented by the values of a set of market factor
information processes. The market filtration is taken to be
generated jointly by the various information processes and by the
default indicator processes of the various bonds. The value of a
discount bond is obtained by taking the discounted expectation of
the value of the default indicator function at the maturity of the
bond, conditional on the information provided by the market
filtration. Explicit expressions are derived for the bond price
processes and the associated default hazard rates. The latter are
not given \textit{a priori} as part of the model but rather are
deduced and shown to be functions of the values of the information
processes. Thus the ``perceived" hazard rates, based on the
available information, determine bond prices, and as perceptions
change so do the prices. In conclusion, explicit expressions are
derived for options on discount bonds, the values of which also
fluctuate in line with the vicissitudes of market sentiment.
\end{abstract}

\maketitle



\section{Credit-risk modelling}

In this paper we consider a simple model for defaultable securities
and, more generally, for a class of financial instruments for which
the cash flows depend on the default times of a set of defaultable
securities. For example, if $\tau$ is the time of default of a firm
that has issued a simple credit-risky zero-coupon bond that matures
at time $T$, then the bond delivers a single cash-flow $H_T$ at $T$,
given by
\begin{eqnarray}
H_T=N\,{\mathds 1}\{\tau>T\} ,
\end{eqnarray}
where $N$ is the principal, and ${\mathds 1}\{\tau>T\}=1$ if
$\tau>T$ and ${\mathds 1}\{\tau>T\}=0$ if $\tau\le T$. By a
``simple'' credit-risky zero-coupon bond we mean the case of an
idealised bond where there is no recovery of the principal if the
firm defaults. If a fixed fraction $R$ of the principal is paid at
time $T$ in the event of default then we have
\begin{equation}
H_T=N{\mathds 1}\{\tau>T\}+RN{\mathds 1}\{\tau\le T\}.
\end{equation}
More realistic models can be developed by introducing random factors
that determine the amount and timing of recovery levels. See, e.g.,
Brody, Hughston \& Macrina (2007, 2010), and Macrina \& Parbhoo
(2010).

As another example, let $\tau_1,\tau_2\ldots,\tau_n$ denote the
default times of a set of $n$ discount bonds, each with maturity
after $T$. Write $\bar{\tau}_1,\bar{\tau}_2,\ldots,\bar{\tau}_n$ for
the ``order statistics'' of the default times. Hence $\bar{\tau}_1$
is the time of the first default (among
$\tau_1,\tau_2,\ldots,\tau_n)$, $\bar{\tau}_2$ is the time of the
second default, and so on. Then a structured product that pays
\begin{eqnarray}
H_T=K\,{\mathds 1}\{\bar{\tau}_k\le T\}
\end{eqnarray}
is a kind of insurance policy that pays $K$ at time $T$ if there
have been $k$ or more defaults by time $T$.

One of the outstanding problems associated with credit-risk
modelling is the following.
What counts in the valuation of credit-risky  products is not necessarily
the ``actual'' or ``objective'' probability of default (even if this can
be meaningfully determined), but rather the ``perceived'' probability
of default.
This can change over time, depending on shifts in market
sentiment and the flow of relevant market information.
How do we go about modelling the dynamics of such products?

\section{Modelling the market filtration}

We introduce a probability space with a measure ${\mathbb Q}$ which,
for simplicity, we take to be the risk-neutral measure. Thus, the
price process of any non-dividend-paying asset, when the price is
expressed in units of a standard money-market account, is a
$\mathbb{Q}$-martingale. We do not assume that the market is
necessarily complete; rather, we merely assume that there is an
established pricing kernel. We assume, again for simplicity, that
the default-free interest-rate term structure is deterministic. Time
$0$ denotes the present, and we write $P_{tT}$ for the price at $t$
of a default-free discount bond that matures at $T$. To ensure
absence of arbitrage, we require that $P_{tT}=P_{0T}/P_{0t}$, where
$\{P_{0t}\}_{0\le t<\infty}$ is the initial term structure. No
attempt will be made in the present investigation to examine the
case of stochastic interest rates: to keep credit-related issues in
the foreground, we suppress considerations relating to the
default-free interest rate term structure. See Rutkowski \& Yu
(2007), Hughston \& Macrina (2009), Brody \& Friedman (2009), and
Macrina \& Parbhoo (2010) for discussions of the stochastic interest
rate case in an information-based setting.

The probability space
comes equipped with a filtration $\{{\mathcal G}_t\}$ which we take
to be the market filtration. Our first objective is to define
$\{{\mathcal G}_t\}$ in such a way that market sentiments concerning
the default times can be modelled explicitly. We let $\tau_1$, $\tau_2$,
\ldots,$\tau_n$ be a collection of
non-negative random times  such that ${\mathbb Q} (\tau_\alpha=0)=0$
and ${\mathbb Q}(\tau_\alpha>t)>0$ for $t>0$ and
$\alpha=1,2,\ldots,n$. We set
\begin{eqnarray}
\tau_\alpha = f_\alpha(X_1,X_2,\ldots,X_N),\qquad
(\alpha=1,2,\ldots,n).
\end{eqnarray}
Here $X_1,X_2,\ldots,X_N$ are $N$ independent, continuous,
real-valued \textit{market factors} that determine the default
times, and $f_\alpha$ for each $\alpha$ is a smooth function of $N$
variables that determines the dependence of $\tau_\alpha$ on the
market factors. We note that if two default times share an
$X$-factor in common, then they will in general be correlated. With
each $\tau_{\alpha}$ we associate a ``survival'' indicator process
${\mathds 1}\{\tau_{\alpha}>t\}$, $t\ge0$, which takes the value
unity until default occurs, at which time it drops to zero.
Additionally, we introduce a set of $N$ information processes
$\{\xi^k_t\}_{t\ge 0}$ in association with the market factors, which
in the present investigation we take to be of the form
\begin{eqnarray}\label{IP}
\xi^k_t=\sigma_k t X_k + B^k_t.
\end{eqnarray}
Here, for each $k$, $\sigma_k$ is a parameter (``information flow
rate'') and $\{B^k_t\}_{t\ge 0}$ is a Brownian motion (``market
noise''). We assume that the $X$-factors and the market noise
processes are all independent of one another.

We take the market filtration $\{{\mathcal G}_t\}_{t\ge0}$ to be
generated jointly by the information processes and the survival
indicator processes. Therefore, we have:
\begin{eqnarray}
{\mathcal G}_t = {\boldsymbol\sigma}
\left[\{\xi_s^k\}_{0\le s\le t}^{k=1,\ldots,N},
{\mathds 1}\{\tau^{\alpha}>s\}^{\alpha=1,\ldots,n}_{0\le s\le t}\right].
\end{eqnarray}
It follows that at $t$ the market knows the information generated up
to $t$ and the history of the indicator processes up to time $t$.
For the purpose of calculations it is useful also to introduce the
filtration $\{{\mathcal F}_t\}_{t\ge0}$ generated by the information
processes:
\begin{eqnarray}\label{F-filtration}
{\mathcal F}_t =  {\boldsymbol\sigma}
\left[\{\xi_s^k\}^{k=1,\ldots,N}_{0\le s\le t}\right]. \label{eq:6}
\end{eqnarray}
Then clearly ${\mathcal F}_t\subset {\mathcal G}_t$. We do not
require as such the notion of a ``background'' filtration for our
theory. Indeed, we can think of the $\xi$'s and the ${\mathds 1}$'s
as providing two related but different types of information about
the $X$'s.

\section{Credit-risky discount bond}

As an example we study in more detail the case $n=1$, $N=1$. We have
a single random market factor $X$ and an associated default time
$\tau=f(X)$. We assume that $X$ is continuous and that $f(X)$ is
monotonic. The market filtration $\{{\mathcal G}_t\}$ is generated
jointly by an information process of the form
\begin{eqnarray}
\xi_t=\sigma t X+B_t,\qquad t\ge 0, \label{eq:8}
\end{eqnarray}
and the indicator process ${\mathds 1}\{\tau>t\}$, $t\ge0$. The
Brownian motion $\{B_t\}_{t>0}$ is taken to be independent of $X$.
The value of a defaultable $T$-maturity discount bond, when there is
no recovery on default, is then given by
\begin{eqnarray}
B_{tT}=P_{tT}\,{\mathbb E}\left[ {\mathds
1}\{\tau>T\}\,\vert\,{\mathcal G}_t\right]
\end{eqnarray}
for $0\le t\le T$. We shall write out an explicit expression for
$B_{tT}$. First, we use the identity (see, e.g., Bielecki, Jeanblanc
\& Rutkowski 2009):
\begin{eqnarray}
{\mathbb E}  \left[{\mathds 1}\{\tau>T\}\,\vert\,{\mathcal G}_t\right]=
{\mathds 1}\{\tau>t\} \frac{{\mathbb E}\left[
{\mathds 1}\{\tau>T\}|{\mathcal F}_t \right]} {{\mathbb E}\left[
{\mathds 1}\{\tau>t\}|{\mathcal F}_t \right]},
\end{eqnarray}
where ${\mathcal F}_t$ is as defined in (\ref{eq:6}). It follows
that
\begin{equation}
B_{tT}=P_{tT}{\mathds 1}\{\tau>t\} \frac{{\mathbb E}\left[ {\mathds
1}\{\tau>T\}|{\mathcal F}_t \right]} {{\mathbb E}\left[ {\mathds
1}\{\tau>t\}|{\mathcal F}_t \right]}.
\end{equation}

We note that the information process $\{\xi_t\}$ has the Markov
property with respect to its own filtration. To see this, it
suffices to check that
\begin{eqnarray}
{\mathbb Q}\left( \xi_t\leq x| \xi_s,\xi_{s_1},\xi_{s_2},\ldots,
\xi_{s_k}\right) = {\mathbb Q}\left( \xi_t\leq x|\xi_s\right) \label{eq:10}
\end{eqnarray}
for any collection of times $t\geq s\geq s_1\geq s_2\geq\cdots\geq
s_k>0$. We observe that for $s>s_1>s_2>s_3>0$, the random variables
$\{B_s/s-B_{s_1}/s_1\}$ and $\{B_{s_2}/s_2-B_{s_3}/s_3\}$ are
independent. This follows directly from a calculation of their
covariance. Hence from the relation
\begin{eqnarray}
\frac{\xi_s}{s}-\frac{\xi_{s_1}}{s_1}=\frac{B_s}{s}-
\frac{B_{s_1}}{s_1}
\end{eqnarray}
we conclude that
\begin{eqnarray}
{\mathbb Q}\left( \xi_t\leq x| \xi_s,\xi_{s_1},
\ldots,\xi_{s_k}\right)&=&{\mathbb Q}\left(\xi_t\leq x| \xi_s,
\frac{\xi_s}{s}-\frac{\xi_{s_1}}{s_1}, \frac{\xi_{s_1}}{s_1}-
\frac{ \xi_{s_2}}{s_2}, \ldots, \frac{\xi_{s_{k-1}}}{s_{k-1}} -
\frac{\xi_{s_k}}{s_k}\right)\nonumber \\ && \hspace{-2.5cm}
={\mathbb Q}\left( \xi_t\leq x| \xi_s, \frac{B_s}{s}-
\frac{B_{s_1}}{s_1}, \frac{B_{s_1}}{s_1}-\frac{B_{s_2}}{s_2},
\ldots, \frac{B_{s_{k-1} }}{s_{k-1}}-\frac{B_{s_k}}{s_k}\right).
\end{eqnarray}
However, since $\xi_t$ and $\xi_s$ are independent of $B_s/s-
B_{s_1}/s_1$, $B_{s_1}/s_1-B_{s_2}/s_2$, $\ldots$, $B_{s_{k-1}}
/s_{k-1}-B_{s_k}/s_k$, the result (\ref{eq:10}) follows.

As a consequence of the Markovian property of $\{\xi_t\}$ and the
fact that $X$ is ${\mathcal F}_\infty$-measurable, we therefore
obtain
\begin{eqnarray}
B_{tT} = P_{tT}  {\mathds 1}\{\tau>t\}\, \frac{{\mathbb E}\left[
{\mathds 1}\{\tau>T\}\,\vert\,\xi_t\right]} {{\mathbb E}\left[
{\mathds 1}\{\tau>t\}\,\vert\,\xi_t\right]}
\end{eqnarray}
for the defaultable bond price. Thus we can write
\begin{eqnarray}
B_{tT}=P_{tT}\,{\mathds 1}\{\tau>t\}\,
\frac{\int^{\infty}_{-\infty}{\mathds 1}\{f(x)>T\}\rho_t(x)\rd x}
{\int^{\infty}_{-\infty}{\mathds 1}\{f(x)>t\}\rho_t(x)\rd x}.
\end{eqnarray}
Here
\begin{eqnarray}
\rho_t(x)={\mathbb E}\left[\delta(X-x)\,\vert\,\xi_t\right]
\end{eqnarray}
is the conditional density for $X$ given $\xi_t$, and a
calculation using the Bayes law shows that:
\begin{eqnarray}
\rho_t(x)=\frac{\rho_0(x)\exp\left[\sigma x\xi_t-\frac{1}{2}
\sigma^2 x^2 t\right]}{\int_{-\infty}^{\infty}\rho_0(x)
\exp\left[\sigma x\xi_t-\frac{1}{2}\sigma^2 x^2 t\right]\rd x},
\end{eqnarray}
where $\rho_0(x)$ is the {\it a priori} density of $X$.
Thus for the bond price we obtain:
\begin{eqnarray}\label{bond price process}
B_{tT}=P_{tT}{\mathds
1}\{\tau>t\}\,\frac{\int^{\infty}_{-\infty}\rho_0(x) {\mathds
1}\{f(x)>T\}\exp\left[\sigma x\xi_t-\frac{1}{2}\sigma^2 x^2
t\right]\rd x}{\int^{\infty}_{-\infty}\rho_0(x){\mathds 1}\{f(x)>t\}
\exp\left[\sigma x\xi_t-\frac{1}{2}\sigma^2 x^2 t\right]\rd x}.
\end{eqnarray}
It should be apparent that the value of the bond fluctuates as
$\xi_t$ changes. This reflects the effects of changes in market
sentiment concerning the possibility of default. Indeed, if we
regard $\tau$ as the default time of an obligor that has issued a
number of different bonds (coupon bonds can be regarded as bundles
of zero-coupon bonds), then similar formulae will apply for each of
the various bond issues.

When the default times of two or more distinct obligors depend on a
common market factor, the resulting bond price dynamics are
correlated and so are the default times. The modelling framework
presented here therefore provides a basis for a number of new
constructions in credit-risk management, including explicit
expressions for the volatilities and correlations of credit-risky
bonds.

Our methodology is to consider models for which each independent
$X$-factor has its own information process. Certainly, we can also
consider the situation where there may be two or more distinct
information processes available concerning the same $X$-factor. This
situation is relevant to models with asymmetric information, where
some traders may have access to ``more'' information about a given
market factor than other traders; see Brody, Davis, Friedman \&
Hughston (2008), Brody, Brody, Meister \& Parry (2010), and Brody,
Hughston \& Macrina (2010).

In principle, a variety of different types of information processes
can be considered. We have in (\ref{IP}) used for simplicity what is
perhaps the most elementary type of information process, a Brownian
motion with a random drift. The linearity of the drift term in the
time variable ensures on the one hand that the information process
has the Markov property, and on the other hand, since the Brownian
term grows in magnitude on average like the square-root of time, that the
drift term eventually comes to dominate the noise term, thus
allowing for the release of information concerning the likely time
of default. Information processes based variously on the Brownian bridge
(Brody, Hughston \& Macrina 2007, 2008a), the gamma bridge (Brody,
Hughston \& Macrina 2008b), and the L\'evy bridge (Hoyle, Hughston \&
Macrina 2009, 2010) have been applied to problems in
finance and insurance.

Our work can be viewed in the context of the growing body of
literature on the role of information in finance and its application
to credit risk modelling in particular. No attempt will be made here
at a systematic survey of material in this line. We refer the
reader, for example, to F\"ollmer, Wu, \& Yor (1999), Kusuoka (1999), 
Duffie \& Lando (2001), 
Jarrow \& Protter (2004), \c{C}etin, Jarrow, Protter \& Yildirim (2004),  
Giesecke (2006), Geman, Madan, \& Yor (2007), 
Coculescu, Geman \& Jeanblanc (2008),  Frey \& 
Schmidt (2009), Bielecki, Jeanblanc \& Rutkowski (2009), and works 
cited therein.

\section{Discount bond dynamics}

Further insight into the nature of the model can be gained by
working out the dynamics of the bond price. An application of Ito
calculus gives the following dynamics over the time interval from
$0$ to $T$:
\begin{eqnarray}
\rd B_{tT}=(r_t+h_t)B_{tT}\rd t+\sigma\Sigma_{tT}\,B_{tT}\rd W_t+
B_{t^-T}
\rd{\mathds 1}\{\tau>t\}.
\end{eqnarray}
Here
\begin{eqnarray}
r_t=-\partial_t\ln(P_{0t})
\end{eqnarray}
is the deterministic short rate of interest, and
\begin{eqnarray}
h_t=\frac{{\mathbb E}\left[\delta(f(X)-t)\,\vert\,{\mathcal F}_t\right]}{{\mathbb E}
\left[{\mathds 1} \{f(X)>t\}\,\vert\,{\mathcal F}_t\right]} \label{eq:17}
\end{eqnarray}
is the so-called hazard rate. It should be evident that if $\tau\le
T$ then when the default time is reached the bond price drops to
zero. The defaultable discount bond volatility $\Sigma_{tT}$ is
given by
\begin{eqnarray}
\Sigma_{tT}=\frac{{\mathbb E}\left[{\mathds
1}\{f(X)>T\}X\,\vert\,{\mathcal F}_t\right]} {{\mathbb E}\left[{\mathds
1}\{f(X)>T\}\,\vert\,{\mathcal F}_t\right]}-\frac{{\mathbb E} \left[{\mathds
1}\{f(X)>t\}X\,\vert\,{\mathcal F}_t\right]}{{\mathbb E}\left[{\mathds 1}
\{f(X)>t\} \, \vert\, {\mathcal F}_t\right]}.
\end{eqnarray}
The process $\{W_t\}$ appearing in the dynamics of $\{B_{tT}\}$ is
defined by the relation:
\begin{eqnarray}
W_t = \int^t_0\!\!{\mathds 1}\{f(X)>s\}\left(\rd\xi_s-\sigma
{\mathbb E}\left[X|{\mathcal G}_s\right]\rd s\right). \label{eq:22}
\end{eqnarray}

To deduce the formulae above we define a one-parameter family of
$\{\mathcal{F}_t\}$-adapted processes $\{F_{tu}\}$ by setting
\begin{eqnarray}
F_{tu} = \int^{\infty}_{-\infty}\rho_0(x) {\mathds 1}\{f(x)>u\}
\exp\left[\sigma x\xi_t-\half\sigma^2 x^2 t\right]\rd x.
\end{eqnarray}
Then the bond price can be written in the form
\begin{eqnarray}
B_{tT}=P_{tT}{\mathds 1}\{\tau>t\} \frac{F_{tT}}{F_{tt}}.
\end{eqnarray}
An application of Ito's lemma gives
\begin{eqnarray}
\frac{\rd F_{tT}}{F_{tT}} = \sigma \frac{{\mathbb E}\left[{\mathds
1}\{f(X)>T\}X\,\vert\,{\mathcal F}_t\right]} {{\mathbb E}\left[{\mathds
1}\{f(X)>T\}\,\vert\,{\mathcal F}_t\right]} \, \rd \xi_t
\end{eqnarray}
and
\begin{eqnarray}
\frac{\rd F_{tt}}{F_{tt}} = - \frac{{\mathbb E}\left[\delta(f(X)-t)
\,\vert\,{\mathcal F}_t\right]}{{\mathbb E}\left[{\mathds 1} \{f(X)>t\}\,
\vert\,{\mathcal F}_t\right]}\, \rd t
+ \sigma \frac{{\mathbb E}\left[{\mathds
1}\{f(X)>T\}X\,\vert\,{\mathcal F}_t\right]} {{\mathbb E}\left[{\mathds
1}\{f(X)>T\}\,\vert\,{\mathcal F}_t\right]} \, \rd \xi_t .
\end{eqnarray}
The desired results then follow at once by use of the relation
\begin{equation}
\rd\left(\frac{F_{tT}}{F_{tt}}\right)=\frac{F_{tT}}{F_{tt}}
\left[\frac{\rd F_{tT}}{F_{tT}}-\frac{\rd
F_{tt}}{F_{tt}}+\left(\frac{\rd F_{tt}}{F_{tt}}\right)^2-\frac{\rd
F_{tT}}{F_{tT}}\frac{\rd F_{tt}}{F_{tt}}\right].
\end{equation}
The process $\{W_t\}_{0\leq t<\tau}$ defined by (\ref{eq:22}) is a
$\{{\mathcal G}_t\}$-Brownian motion. This can be seen by use of
L\'evy's characterisation. Specifically, we have $\rd W_t^2=\rd t^2$
and ${\mathbb E}[W_u\,\vert\,\mathcal{G}_t]=W_t$. To see that
$\{W_t\}_{0\le t<\tau}$ is a $\{{\mathcal G}_t\}$-martingale we
observe that
\begin{eqnarray}
W_u&=&\int^u_0{\mathds
1}\left\{\tau>s\right\}\left(\rd\xi_s-\sigma\mathbb{E}\left[X\,\vert\,
\mathcal{G}_s\right]\rd
s\right)\nonumber,\\
&=&W_t+\int^u_t{\mathds
1}\{\tau>s\}\left(\rd\xi_s-\sigma\mathbb{E}\left[X\,\vert\,\mathcal{G}_s\right]\rd
s\right),
\end{eqnarray}
and hence
\begin{equation}
\mathbb{E}\left[W_u\,\vert\,\mathcal{G}_t\right]=W_t+\mathbb{E}
\left[\int^u_t{\mathds
1}\left(\rd\xi_s-\sigma\mathbb{E}[X\,\vert\,\mathcal{G}_s]\rd
s\right)\bigg\vert\,\mathcal{G}_t\right].
\end{equation}
Then by inserting $\rd\xi_s=\sigma X\rd s+\rd B_s$ and using the
tower property we find that the terms involving $X$ cancel and we
are left with
\begin{eqnarray}
\mathbb{E}\left[W_u\,\vert\,\mathcal{G}_t\right]&=&W_t+\mathbb{E}
\left[\int^u_t{\mathds
1}\{\tau>s\}\rd B_s\,\bigg\vert\,\mathcal{G}_t\right],\nonumber\\
&=&W_t+\mathbb{E}\left[\mathbb{E}\left[\int^u_t{\mathds
1}\{\tau>s\}\rd B_s\,\bigg\vert\,\sigma\left\{X,\mathcal{G}_t\right\}\right]\,
\bigg\vert\,\mathcal{G}_t\right],\nonumber\\
&=&W_t.
\end{eqnarray}

The Brownian motion that ``drives" the defaultable bond is not
adapted to a pre-specified background filtration. Rather, it is
directly associated with information about the factors determining
default. In this respect, the information-based approach is closer
in spirit to a structural model, even though it retains the economy
of a reduced-form model.

\section{Hazard rates and forward hazard rates}

Let us now examine more closely properties of the intensity process
$\{h_t\}$ given by the expression (\ref{eq:17}). We remark first
that the intensity at time $t$ is a function of $\xi_{t}$. This shows that in
the present model the default intensity is determined by
``market perceptions.'' Our model can thus be characterised as follows:

The market does not know the ``true" default intensity; rather,
from the information available to the market a kind of ``best
estimate" for the default intensity is used for the pricing of
bonds---but the market is ``aware" of the fact that this estimate
is based on perceptions, and hence as the perceptions change,
so will the estimate, and so will the bond prices. Thus, in the
present model (and unlike the majority of credit models hitherto
proposed) there is no need for ``fundamental changes'' in the
state of the obligor, or the underlying economic environment,
as the basis for improvement or deterioration of credit quality:
it suffices simply that the information about the credit quality
should change---whether or not this information is actually
representative of the true state of affairs.

In the present example we can obtain a more explicit expression for
the intensity by transforming the variables as follows. Since $f$ is
invertible, we can introduce the inverse function $\phi(\tau)=
f^{-1}(\tau)=X$ and write
\begin{eqnarray}
\xi_t = \sigma t \phi(\tau) + B_t \label{eq:33x}
\end{eqnarray}
for the information process. As before, we assume that the Brownian
motion $\{B_t\}_{t\ge 0}$ is independent of the default time $\tau$.
Writing $p(u)$ for the \textit{a priori} density of the random
variable $\tau=f(X)$ we then deduce from (\ref{eq:17}) that the
hazard process is given by have the expression
\begin{eqnarray}
h_t = \frac{ p(t)\, \re^{\sigma \phi(t) \xi_t-\frac{1}{2}
\sigma^2 \phi^2(t)t} }{\int_t^\infty p(u)\, \re^{\sigma \phi(u)
\xi_t - \frac{1}{2}\sigma^2 \phi^2(u) t}\rd u}. \label{eq:9}
\end{eqnarray}
This expression manifestly reminds us the fact that $\{h_t\}$ is a
function of the information $\xi_t$, and thus its value moves up and
down according to the market perception of the timing of default.

In the present context it is also natural to consider the
forward hazard rate defined by the expression
\begin{eqnarray}
h_{tu}=\frac{{\mathbb E}\left[\delta(f(X)-u)\,\vert\,{\mathcal F}_t\right]}
{{\mathbb E}\left[{\mathds 1} \{f(X)>u\}\,\vert\,{\mathcal F}_t\right]}.
\end{eqnarray}
We observe that $h_{tu}\rd u$ represents the \textit{a posteriori}
probability of default in the infinitesimal interval $[u,u+\rd u]$,
conditional on no default until time $u$. More explicitly, we have
\begin{eqnarray}
h_{tu} = \frac{ p(u)\, \re^{\sigma \phi(u) \xi_t-\frac{1}{2}
\sigma^2 \phi^2(u) t}}{\int_u^\infty p(v)\, \re^{\sigma \phi(v)
\xi_t - \frac{1}{2}\sigma^2 \phi^2(v) t}\rd v} \label{eq:10x}
\end{eqnarray}
for the forward hazard rate.

\begin{figure}[th]
  \includegraphics[scale=0.47]{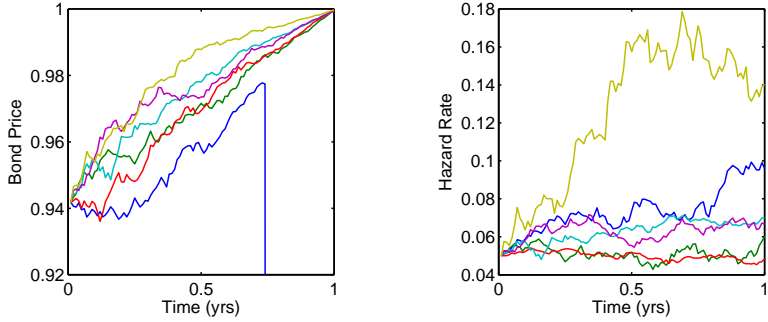}
  \vspace{-0.0cm}
  \caption{Sample paths of the defaultable discount
  bond and the associated hazard rate. We choose
  $\phi(t)=\re^{-0.025t}$,
  and the initial term structure is assumed to be flat so that
  $P_{0t}=\re^{-0.02t}$. The information flow rate parameter is set to be
  $\sigma=0.3$, and the bond maturity is one year.
  \label{fig:1}
  }
\end{figure}

We remark that it is a straightforward matter to simulate the
dynamics of the bond price and the associated hazard rates by Monte
Carlo methods. First we simulate a value for $X$ by use of the
\textit{a priori} density $\rho_0(x)$. From this we deduce the
corresponding value of $\tau$. Then we simulate an independent
Brownian motion $\{B_t\}$, and thereby also the information process
$\{\xi_t\}$. Putting these ingredients together we have a simulation
for the bond price and the associated hazard rate. In
figure~\ref{fig:1} we sketch several sample paths resulting from
such a simulation study.

\section{Options on defaultable bonds}

We consider the problem of pricing an option on a defaultable bond
with bond maturity $T$. Let $K$ be the option strike price and let
$t~(<T)$ be the option maturity. The payoff of the option is then
$(B_{tT}-K)^+$. Let us write the bond price at in the form
\begin{equation}\label{altB}
B_{tT}={\mathds 1}\{\tau>t\}B(t,\xi_t),
\end{equation}
where the function $B(t,y)$ is defined by
\begin{eqnarray}
B(t,y)=P_{tT}\, \frac{\int^{\infty}_{T}p(u) \exp\left[
\sigma \phi(u)y-\frac{1}{2}\sigma^2 \phi^2(u) t\right]\rd u}
{\int^{\infty}_{t}p(u)\exp\left[\sigma \phi(u)
y-\frac{1}{2}\sigma^2 \phi^2(u) t\right] \rd u}.
\end{eqnarray}
We make note of the identity
\begin{equation}
\big({\mathds 1}\{\tau>t\}B(t,\xi_t)-K\big)^+
={\mathds 1}\{\tau>t\}\big(B(t,\xi_t)-K\big)^+,
\end{equation}
satisfied by the option payoff. The price of the option is thus
given by
\begin{eqnarray}
C_0 = P_{0t} \,
{\mathbb E}\Big[ {\mathds 1}\{\tau>t\}\big(B(t,\xi_t)-K\big)^+ \Big] .
\label{eq:46x}
\end{eqnarray}
We find, in particular, that the option payoff is a function of the
random variables $\tau$ and $\xi_t$. To determine the expectation
(\ref{eq:46x}) we need therefore to work out the joint density of
$\tau$ and $\xi_t$, defined by
\begin{eqnarray}
\rho(u,y) = {\mathbb E}\left[ \delta(\tau-u)\delta(\xi_t-y)\right] =
- \frac{\rd}{\rd u}\, {\mathbb E}\left[ {\mathds 1}\{\tau>u\}
\delta(\xi_t-y)\right] . \label{eq:47}
\end{eqnarray}
Note that the expression
\begin{eqnarray}
A_{0}(u,y) = P_{0t}\, {\mathbb E}\left[ {\mathds 1}\{\tau>u\}
\delta(\xi_t-y)\right]
\end{eqnarray}
appearing on the right side of (\ref{eq:47}), with an additional
discounting factor, can be regarded as representing the price of a
``defaultable Arrow-Debreu security'' based on the value at time $t$
of the information process.

\begin{figure}[th]
\begin{center}\hspace{-0.0cm}
  \includegraphics[scale=.65]{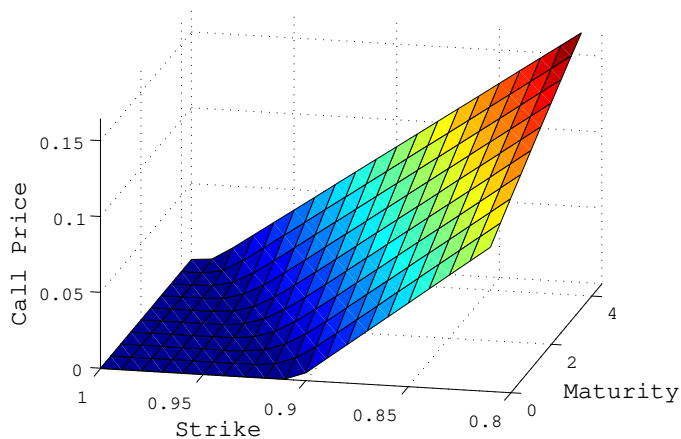}
  \vspace{-0.0cm}
  \caption{Price of a call option on a defaultable discount
  bond. The bond maturity is five years. The information-adjusting
  function is set to be $\phi(t)=\re^{-0.05t}$, and  the initial term
  structure is assumed to be flat so that $P_{0t}=\re^{-0.02t}$. The
  information flow rate is set to be $\sigma=0.25$.
  \label{fig:2}
  }
\end{center}
\end{figure}

To work out the expectation appearing
here we shall use the Fourier representation for the delta function:
\begin{equation}
\delta(\xi_t-y)=\frac{1}{2\pi}\int_{-\infty}^\infty \exp\left(-\ri
y\lambda+\ri \xi_t\lambda\right)\rd\lambda,
\end{equation}
which of course has to be interpreted in an appropriate way with
respect to integration against a class of test functions. We have
\begin{equation}\label{E1}
\mathbb{E}\left[{\mathds 1}\{\tau>u\}\delta(\xi_t-y)\right] =
\frac{1}{2\pi}\int_{-\infty}^\infty \re^{-{\rm i}y\lambda}
\mathbb{E} \left[{\mathds 1}\{\tau>u\}\re^{{\rm
i}\xi_t\lambda}\right] \rd\lambda.
\end{equation}
The expectation appearing in the integrand is given by
\begin{equation}\label{E2}
\mathbb{E}\left[{\mathds 1}\{\tau>u\}\re^{{\rm i}\xi_t\lambda}
\right]=\int_{-\infty}^\infty {\mathds 1}\{x>u\}\, p(x)
\exp\left(\ri\sigma\lambda t \phi(x) - \half \lambda^2 t\right)\rd x,
\end{equation}
where we have made use of the fact that the random variables $\tau$
and $B_t$ appearing in the definition of the information process
(\ref{eq:33x}) are independent. We insert this intermediate result
in (\ref{E1}) and rearrange terms to obtain
\begin{equation}
\mathbb{E}\left[{\mathds 1}\{\tau>u\}\delta(\xi_t-y)\right] =
\frac{1}{2\pi} \int_{-\infty}^\infty\!\!\!{\mathds 1}\{x>u\}\, p(x)
\!\int_{-\infty}^\infty\!\! \re^{-{\rm i}y\lambda + {\rm i}\sigma\lambda t
\phi(x) - \frac{1}{2}\lambda^2 t}\rd\lambda \rd x.
\end{equation}
Performing the Gaussian integration associated with the $\lambda$
variable we deduce that the price of the defaultable Arrow-Debreu
security is
\begin{equation}
A_{0}(u,y)=\frac{P_{0t}}{\sqrt{2\pi\,t}}\int_{-\infty}^\infty
{\mathds 1}\{x>u\} p(x)\exp\left[-\frac{(\sigma t \phi(x)-y)^2}
{2t}\right]\rd x. \label{eq:48}
\end{equation}
For the calculation of the price of a call option written on a
defaultable discount bond, it is convenient to rewrite (\ref{eq:48})
in the following form:
\begin{equation}\label{AD}
A_{0}(y)=\frac{P_{0t}}{\sqrt{2\pi\,t}}\exp\left(-
\frac{y^2}{2t}\right)\int_{u}^\infty p(x)\exp\left(\sigma \phi(x)
y-\half\,\sigma^2 \phi^2(x) t\right)\rd x.
\end{equation}
It follows that for the joint density function we have
\begin{eqnarray}
\rho(u,y) = \frac{1}{\sqrt{2\pi t}}\, p(u)\, \re^{-\frac{1}{2t}y^2}
\, \re^{ \sigma \phi(u) y-\frac{1}{2}\,\sigma^2 \phi^2(u) t}.
\end{eqnarray}
The price of the call option can therefore be worked out as follows:
\begin{eqnarray}
C_0 &=& P_{0t} \int_{-\infty}^\infty \rd u \int_{-\infty}^\infty \rd y \,
\rho(u,y) {\mathds 1}\{u>t\} (B(t,y)-K)^+  \nonumber \\
&=& \frac{P_{0t}}{\sqrt{2\pi t}} \int_{-\infty}^\infty \!\!\! \rd y \,
\re^{-\frac{1}{2t}y^2} (B(t,y)-K)^+ \left[ \int_t^\infty p(u)
\re^{\sigma \phi(u) y-\frac{1}{2}\sigma^2 \phi^2(u) t} \rd u \right] .
\,\,\,\,
\end{eqnarray}
We notice that the term inside the square brackets is identical to
the denominator of the expression for $B(t,y)$. Therefore, we have
\begin{eqnarray}
C_0 &=& \frac{P_{0t}}{\sqrt{2\pi t}} \int_{-\infty}^\infty \!\!\!
\rd y \, \re^{-\frac{1}{2t}y^2} \left(  P_{tT} \int_T^\infty p(u)
\re^{\sigma \phi(u) y-\frac{1}{2}\sigma^2 \phi^2(u) t} \rd u \right.
\nonumber \\ && \qquad \qquad \qquad \qquad \qquad \left. - K
\int_t^\infty p(u) \re^{\sigma \phi(u) y-\frac{1}{2}\sigma^2
\phi^2(u) t} \rd u \right)^+. \label{eq:38}
\end{eqnarray}
An analysis similar to the one carried out in Brody \& Friedman
(2009) shows the following result: Provided that $\phi(u)$ is a
decreasing function there exists a unique critical value $y^*$ for
$y$ such that
\begin{eqnarray}
 P_{tT} \int_T^\infty p(u)
\re^{\sigma \phi(u) y-\frac{1}{2}\sigma^2 \phi^2(u) t} \rd u - K
\int_t^\infty p(u) \re^{\sigma \phi(u) y-\frac{1}{2}\sigma^2
\phi^2(u) t} \rd u > 0
\end{eqnarray}
if $y<y^*$. On the other hand, if $\phi(u)$ is increasing, then there
is likewise a unique critical value $y^\dagger$ of $y$ such that
for $y>y^\dagger$ we have
\begin{eqnarray}
 P_{tT} \int_T^\infty p(u)
\re^{\sigma \phi(u) y-\frac{1}{2}\sigma^2 \phi^2(u) t} \rd u - K
\int_t^\infty p(u) \re^{\sigma \phi(u) y-\frac{1}{2}\sigma^2
\phi^2(u) t} \rd u > 0.
\end{eqnarray}
Therefore, we can perform the $y$-integration in (\ref{eq:38}) to
obtain the value
\begin{eqnarray}
C_0&=&P_{0T} \int_T^\infty p(u)\, N\left( \frac{y^*-\sigma\phi(u)t}
{\sqrt{t}}\right) \rd u \nonumber \\ && \qquad \qquad
- P_{0t}K \int_t^\infty p(u) \, N\left( \frac{y^*-
\sigma\phi(u)t}{\sqrt{t}}\right) \rd u
\end{eqnarray}
when $\phi(u)$ is decreasing. If $\phi(u)$ is increasing, we have
\begin{eqnarray}
C_0&=&P_{0T} \int_T^\infty p(u) \, N\left( \frac{\sigma\phi(u)t
-y^\dagger}{\sqrt{t}}\right) \rd u \nonumber \\ && \qquad \qquad
- P_{0t}K \int_t^\infty p(u)\, N\left( \frac{\sigma\phi(u)t-y^\dagger}
{\sqrt{t}}\right) \rd u.
\end{eqnarray}
The critical values $y_\phi^*(t,T,K,\sigma)$ and
$y_\phi^\dagger(t,T,K, \sigma)$ can be determined numerically to
value the prices of call options. An example of the call price as a
function of its strike and maturity, when $\phi(u)$ is decreasing,
is shown in figure~\ref{fig:2}. If the function $\phi(u)$ is not
monotonic, then there is in general more than one critical value of
$y$ for which the argument of the max function in (\ref{eq:38}) is
positive. Therefore, in this case there will be more terms in the
option-valuation formula.

The case represented by a simple discount bond is merely an example
and as such cannot be taken too seriously as a realistic model. Nevertheless 
it is interesting that modelling the information available about the
default time leads to a dynamical model for the bond price, in which
the Brownian fluctuations driving the price process arise in a
natural way as the innovations associated with the flow of
information to the market concerning the default time. Thus no
``background filtration" is required for the analysis of default in
models here proposed. The information flow-rate parameter $\sigma$
is not directly observable, but rather can be backed out from
option-price data on an ``implied" basis. The extension of the present 
investigation, which
can be regarded as a synthesis of the ``density" approach to
interest-rate modelling proposed in Brody \& Hughston (2001, 2002)
and the information-based asset pricing framework developed in
Brody, Hughston \& Macrina (2007, 2008a, 2008b) and Macrina (2006),
to multiple asset situations with portfolio credit risk will be
pursued elsewhere.

\section*{Acknowledgments}
The authors are grateful to seminar participants at the Workshop on
Incomplete Information on Mathematical Finance, Chemnitz (June
2009), the Kyoto Workshop on Mathematical Finance and Related Topics
in Economics and Engineering (August 2009), Quant Congress Europe,
London (November 2009), King's College London (December 2009),
Hitotsubashi University, Tokyo (February 2010), and Hiroshima
University (February 2010) for helpful comments and suggestions.
Part of this work was carried out while LPH was visiting the Aspen
Center for Physics (September 2009), and Kyoto University (August
and October 2009). DCB and LPH thank HSBC, Lloyds TSB, Shell
International, and Yahoo Japan for research support.

\end{document}